**Observation of magnon torques mediated by orbital hybridization at the light metal/antiferromagnetic insulator interface**

Yuchen Pu,[1] Guoyi Shi,[1] Hua Bai,[1] Xinhou Chen,[1] Chenhui Zhang,[1] Zhaohui Li,[1] Mehrdad Elyasi,[2] and Hyunsoo Yang[1]

[1]Department of Electrical and Computer Engineering, National University of Singapore, Singapore 117583, Singapore
[2]Advanced Institute for Materials Research, Tohoku University, Sendai 980-8577, Japan

**ABSTRACT**. Magnon torques, which can operate without involving moving electrons, could circumvent the Joule heating issue. In conventional magnon torque systems, the spin source layer with strong spin-orbit coupling is utilized to inject magnons, and the efficiency is limited by the inherent spin Hall conductivity of the spin source layer. In this work, we observe magnon torques in the Cr/NiO/ferromagnet heterostructure with the effective spin Hall conductivity of $2.45 \times 10^5$ $\hbar/2e$ $\Omega^{-1}$ m$^{-1}$, twice that of the best conventional magnon torque system. We demonstrate the magnon-torque-driven switching of a perpendicularly magnetized CoFeB layer at room temperature, with a switching power consumption density of 0.136 mW μm$^{-2}$. We find that the magnon torque originates from the orbital hybridization and interfacial inversion symmetry breaking at the Cr/NiO interface. Our findings not only significantly enhance the efficiency of magnon torques, but also provide key insights into the fundamental mechanisms of magnon injections.

## I. INTRODUCTION.

Magnon quasiparticles, in contrast to electrons, can transport spin angular momentum in insulating systems, offering a promising route to circumvent Joule heating issues [1]. Furthermore, magnons propagation in insulators can reach micrometer-scale distances [2-4] at ultrafast velocities [5]. Benefiting from the above attributes, magnons have emerged as a tool for the manipulation of magnetic moments, with magnon currents exerting magnon torques in the adjacent ferromagnet (FM) [6]. So far, in magnon torque systems, magnons are injected into the antiferromagnetic insulator (AFMI) via spin currents in the spin source (SS) layer, such as Bi$_2$Se$_3$ [6], Pt [7], Hf [8], SrIrO$_3$ [9], Bi$_2$Te$_3$ [10], and WTe$_2$ [11]. However, despite the bulk spin Hall conductivity (SHC) of the SS layers such as Pt, which can reach 2.5 × 10$^5$ $\hbar/2e$ $\Omega^{-1}$ m$^{-1}$, the torque efficiency of magnon-torque-driven magnetization switching remains relatively low due to angular momentum loss in magnon injection and propagation.

It has been argued that the orbital Hall conductivity (OHC) combined with spin-orbit coupling (SOC) leads to the SHC of heavy metals like Pt and W [12,13]. The OHC is predicted to be even larger for light metals despite their small SHC [14,15]. Therefore, utilizing the OHC for magnon injections holds promise for enhancing the efficiency of magnon-torque-driven switching. Efficient magnon injection and detection have recently been demonstrated in the oxidized Cu (CuO$_x$)/Pt/YIG heterostructures [16]. In this system, orbital currents generated in the CuO$_x$ layer are converted into spin currents in the adjacent Pt layer with strong SOC [17,18], and then inject magnons into the YIG layer. Notably, even without the Pt layer, interconversion between orbital currents and the orbital component of magnetization dynamics has been observed in Bi-doped YIG [19] and α-Fe$_2$O$_3$ [20]. However, the direct manipulation of ferromagnetic magnetization by magnon torques driven by the orbital Hall effect (OHE) [21-24] or orbital Rashba-Edelstein effect (OREE) [17,25,26] has not been demonstrated. In addition, the observation of magnon injection in the absence of a Pt layer, using a material with weak SOC, can offer valuable insight into the fundamental mechanisms of magnon injection.

In this letter, we report the observation of magnon torques in a sandwich heterostructure composed of Cr/NiO/FM. The effective spin Hall conductivity, which is defined as the magnon torque efficiency divided by the resistivity of the nonmagnet (NM), is significantly enhanced compared to conventional magnon torque systems. Furthermore, we demonstrate magnon-torque-driven switching of a perpendicularly magnetized CoFeB layer at room temperature, achieving both higher switching efficiency and lower power consumption than previously reported magnon torque systems. Given the weak SOC of bulk in Cr, the observed magnons cannot be injected via the spin Hall effect. In addition, we provide experimental evidence that the magnon torque is generated by orbital hybridization and interfacial inversion symmetry breaking at the Cr/NiO interface.

## II. RESULTS

In conventional magnon torque systems, as illustrated in Fig. 1(a), an electron current ($J_e$) flowing through the SS layer generates a transverse spin current ($J_S$) via the spin Hall effect. The spin-polarized electrons transfer their spin angular momentum to the localized magnetic moments in the AFMI, exciting magnons. The magnon current propagates through the AFMI and exerts magnon torques on the adjacent FM layer, inducing magnetization switching. In contrast, when the SS layer is replaced with a light metal (LM), as shown in Fig. 1(b), the $J_e$ can generate an orbital angular momentum accumulation via the OREE due to interfacial symmetry breaking induced by orbital hybridization. While the orbital accumulation cannot directly interact with magnetic moments through exchange [27], it can be converted into a spin accumulation, which in turn excites magnons into the AFMI. We select chromium (Cr) as the LM and NiO as the AFMI, since Cr exhibits weak SOC but a large OHC. The experimental details are listed in Sec. I, Supplemental Material [28]. The thickness of Cr is chosen to be 6 nm, corresponding to its reported orbital diffusion length [29]. The NiO is polycrystalline and exhibits acceptable surface roughness, as discussed in Sec. II of Supplemental Material [28].

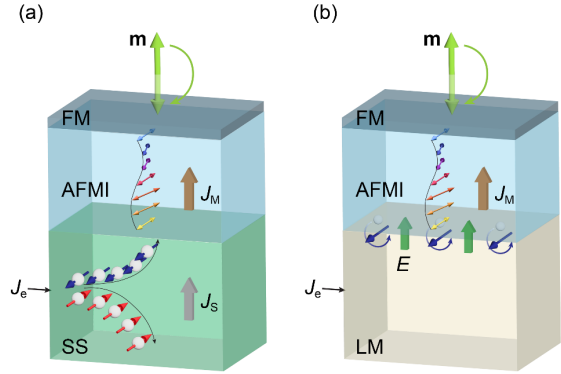

FIG 1. (a) Magnon-torque-driven magnetization switching in conventional trilayers consisting of a spin source (SS), an antiferromagnetic insulator (AFMI), and a ferromagnet (FM). $J_e$ represents the electron current, $J_S$ represents the spin current, and $J_M$ represents the magnon current. The blue and red arrows in the SS layer represent the spin angular momentum. (b) Spin and orbital angular momentum are accumulated at the light metal (LM)/AFMI interface via the orbital Rashba-Edelstein effect (OREE), inducing magnon-torque-driven magnetization switching. $E$ represents the electric field generated at the interface. The blue arrow at the interface between the SS and AFMI layers represents the spin angular momentum.

We utilize the spin-torque ferromagnetic resonance (ST-FMR) technique [30,31] to quantify the magnon torque. Figure 2(a) depicts the representative ST-FMR spectrum ($V_{mix}$) for the Cr (6 nm)/Py (6 nm) sample. The $V_{mix}$ can be decomposed into symmetric ($V_S F_S$) and antisymmetric ($V_A F_A$) Lorentzian components. $V_S F_S$ originates from the in-plane anti-damping torque, while $V_A F_A$ arises from the field-like and Oersted-field torques. According to the Py thickness independence of the torque efficiency (see Sec. III of Supplementary Material [28]), the field-like torque is negligible in our samples, confirming that the $V_S/V_A$ ratio method can be reliably applied to extract the damping-like torque efficiency [31,32]. For the Cr/Py sample, the amplitude of $V_S$ is much smaller than $V_A$ ($V_S/V_A = 0.05$), indicating a weak torque exerting on the Py layer. This observation is consistent with previous reports that Py is not an optimal ferromagnet for orbital torque detection compared to Ni [27,33,34]. In contrast, as shown in Fig. 2(b), the amplitude of $V_S$ for the Cr (6 nm)/NiO (25 nm)/Py (6 nm) sample is comparable to $V_A$ ($V_S/V_A = 0.72$), indicating that the Py layer absorbs a strong torque. Since the NiO layer is insulating and suppresses free electron-mediated spin currents, the observed torque must be exerted by magnons propagating in the NiO layer, indicating that magnons are injected via the Cr layer with weak SOC.

To validate the results obtained from ST-FMR, we conduct in-plane second harmonic Hall (SHH) measurements to extract the torque efficiency ($\theta_T$) (see Sec. IV of Supplemental Material [28]). When an alternative current is applied and an external magnetic field ($\mu_0 H_{ext}$) is rotated in-plane at $\varphi$ relative to the current direction, the second harmonic resistances can be expressed as [35],

$$R_{xy}^{2\omega}(\varphi) = R_{DL+ANE}^{2\omega} \cos(\varphi) + R_{FL}^{2\omega} \cos(\varphi)\cos(2\varphi) \quad (1)$$

where $R_{DL+ANE}^{2\omega}$ is related to the damping-like torque and anomalous Nernst effect, and $R_{FL}^{2\omega}$ is related to the field-like torque.

As shown in Fig. 2(c), we extract the $R_{DL+ANE}^{2\omega}$ values by fitting the second harmonic resistance $R_{xy}^{2\omega}(\varphi)$ of the Cr/Py sample using Eq. (1), which can be expressed as

$$R_{DL+ANE}^{2\omega} = -\frac{1}{2}\left(\frac{R_{AHE}H_{DL}}{H_{ext}+H_k} + I_\omega \alpha \nabla T\right) \quad (2)$$

$$\theta_T = 2eH_{DL}M_s t_{FM}/(\hbar J_c) \quad (3)$$

where $H_{DL}$ is the effective field due to damping-like torque, $H_k$ is the anisotropic field, $\nabla T$ is the periodic temperature gradient induced by currents, $\alpha$ is the Nernst coefficient, $M_s$ is the saturation magnetization of Py, $t_{FM}$ is the thickness of the Py layer, and $J_c$ is the current density in the Cr layer. In Fig. 2(d), the $R_{DL+ANE}^{2\omega}/R_{AHE}$ of the Cr/Py sample is plotted as a linear function of $1/\mu_0(H_{ext}+H_k)$, where the slope of the linear fit corresponds to the strength of the effective field due to damping-like torque, as described in Eq. (3). The nearly flat slope indicates a weak torque, with a corresponding $\theta_T$ of 0.021. However, as shown in Fig. 2(e) and (f), the steeper slope observed for the Cr/NiO/Py samples indicates a stronger torque, with an estimated $\theta_T$ of 0.101.

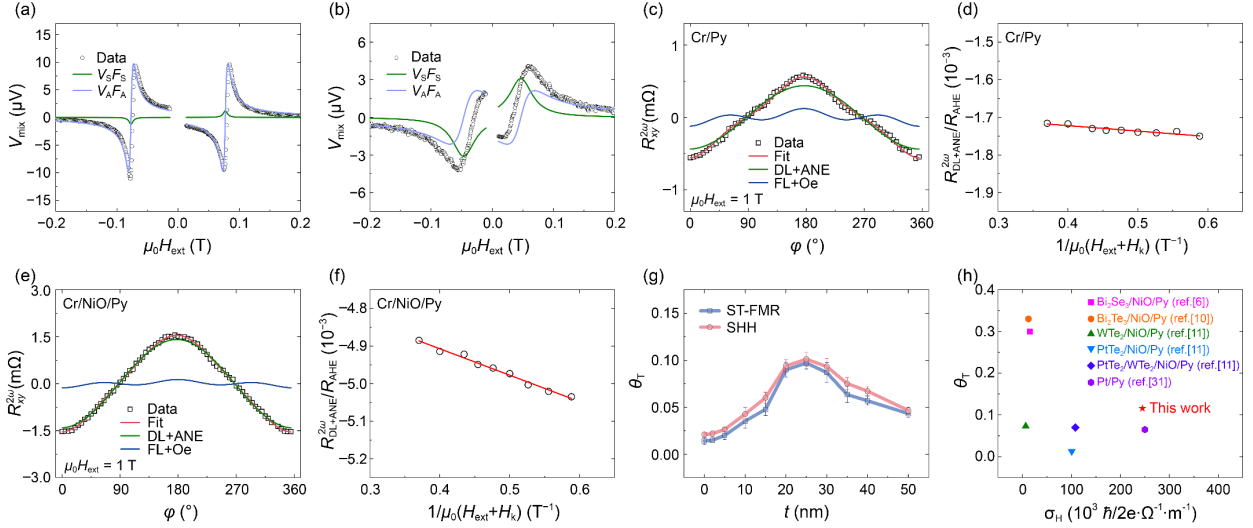

FIG. 2. ST-FMR spectra of (a) Cr (6 nm)/Py and (b) Cr (6 nm)/NiO (25 nm)/Py. The green and purple lines are the symmetry ($V_S F_S$) and anti-symmetry ($V_A F_A$) Lorentzian contributions, respectively. The second harmonic Hall signal ($R_{xy}^{2\omega}$) for (c) Cr/Py and (e) Cr/NiO/Py as a function of $\varphi$. $R_{DL+ANE}^{2\omega}/R_{AHE}$ for (d) Cr/Py and (f) Cr/NiO/Py as a function of the external magnetic field and anisotropic field ($1/\mu_0(H_{ext} + H_k)$). (g) NiO thickness ($t$) dependence of the torque efficiency ($\theta_T$) in Cr (6 nm)/NiO ($t$)/Py. (h) Comparison of the effective spin Hall conductivity ($\sigma_H$) and $\theta_T$.

As shown in Fig. 2(g), the $\theta_T$ values extracted from ST-FMR measurements are consistent with these obtained from in-plane SHH measurements. At $t = 0$ nm, the negligible $\theta_T$ of 0.014 suggests that orbital currents cannot directly induce torques without conversion. As $t$ increases, the amplitudes of $\theta_T$ progressively rise, reaching a maximum value of 0.097 at $t = 25$ nm. This peak corresponds to the emergence of antiferromagnetic ordering in the NiO layer [36] (see Sec. VIII of Supplemental Material [28]), which is consistent with previous observations in conventional magnon torque systems [6,10,11]. For $t > 25$ nm, $\theta_T$ experiences an exponential decay due to dissipation induced by magnon-phonon and magnon-magnon interactions within the NiO layer [37]. The decay is fitted by an exponential decay function,

$$\theta_T = \theta_p \exp\left(-(t - t_p)/l_m\right) \quad (4)$$

where $t_p = 25$ nm is the NiO thickness at peak torque efficiency, $\theta_p = 0.097$ is the corresponding peak value, and $l_m$ represents the magnon diffusion length. From the fit, $l_m$ is determined to be 28.9 nm, which is comparable to previous reports on NiO-based magnon torque systems [6,10]. In addition, The THz emission amplitude exhibits a thickness dependence similar to that of $\theta_T$, verifying the observation of magnon torque. [5,38,39] (See Sec. VI of Supplemental Material [28]). Strong magnon torque is also observed in the Cr/NiO/CoFeB system, indicating that the effect is independent of the choice of ferromagnets (see Sec. XII of Supplemental Material [28]).

The effective spin Hall conductivities ($\sigma_H = \theta_T/\rho_{NM}$, where $\rho_{NM}$ is the resistivity of the NM layer) are summarized in Fig. 2(h). For the Cr/NiO/Py sample, $\sigma_H$ reaches $2.45 \times 10^5$ $\hbar/2e$ $\Omega^{-1}$ m$^{-1}$, surpassing the best conventional magnon torque system, PtTe$_2$/WTe$_2$/NiO/Py ($1.08 \times 10^5$ $\hbar/2e$ $\Omega^{-1}$ m$^{-1}$) [11], comparable with typical electron-mediated spin source materials such as Pt ($2.5 \times 10^5$ $\hbar/2e$ $\Omega^{-1}$ m$^{-1}$) [31].

We subsequently utilize magnon torques to achieve magnetization switching of a FM with perpendicular magnetization anisotropy (PMA). A Ti (2 nm)/Co$_{20}$Fe$_{60}$B$_{20}$ (1 nm)/MgO (2 nm)/TaO$_x$ (1.5 nm) stack is deposited on top of Cr (6 nm)/NiO (25 nm), as illustrated in Fig. 3(a). Ti is selected as the buffer layer due to its long spin diffusion length and its ability to enhance the PMA of the CoFeB layer [10]. The 2 nm Ti layer is too thin to generate a sufficient orbital current, and cannot contribute to the magnetization switching (see Sec. X of Supplemental Material [28]). For comparison, a reference sample Pt (5 nm)/CoFeB, which exhibits the highest spin torque efficiency at this thickness [40], is prepared. The PMA of the CoFeB layer is confirmed by the square-shaped anomalous Hall loop shown in Fig. 3(b). Notably, the comparable coercivity observed in the Cr/NiO/CoFeB and Pt/CoFeB samples indicates similar magnetic properties of the CoFeB layer, allowing for the use of critical current density ($J_c$) to compare the switching efficiency. As shown in Fig. 3(c), clear switching windows are observed with an external magnetic field ($\mu_0 H_x$) applied along the current direction to break the symmetry [31,41,42]. The switching ratio, which is defined as the ratio of switching resistance ($R_{SW}$) to the anomalous Hall resistance ($R_{AHE}$), reaches 100 %. This switching ratio is higher than that for the $Bi_2Te_3$ (8 nm)/NiO (25 nm)/CoFeB system with a switching ratio of 70 % [10], highlighting the enhanced switching efficiency in the Cr/NiO/CoFeB system. The switching is clockwise at +10 mT and anticlockwise at −10 mT, which is similar to typical spin-torque PMA switching induced by Pt [43]. Figure 3(d) presents the total switching current ($I_c$) and $J_c$ within the NM layer under various $\mu_0H_x$, showing that $J_c$ for the Cr/NiO/CoFeB and Pt/CoFeB samples slightly decreases with increasing $\mu_0H_x$. This behavior is similar to the case of electron-mediated spin-torque switching [42]. Notably, the $I_c$ and $J_c$ values for the Cr/NiO/CoFeB sample are comparable to those of the Pt/CoFeB sample, indicating that the large effective spin Hall conductivity in the Cr/NiO/FM system enables switching efficiencies comparable with electron-mediated spin-torque switching. In addition, the influence of current shunting through the CoFeB layer is excluded by demonstrating the magnetization switching of the CoFeB dot (see Sec. XIII of Supplemental Material [28]).

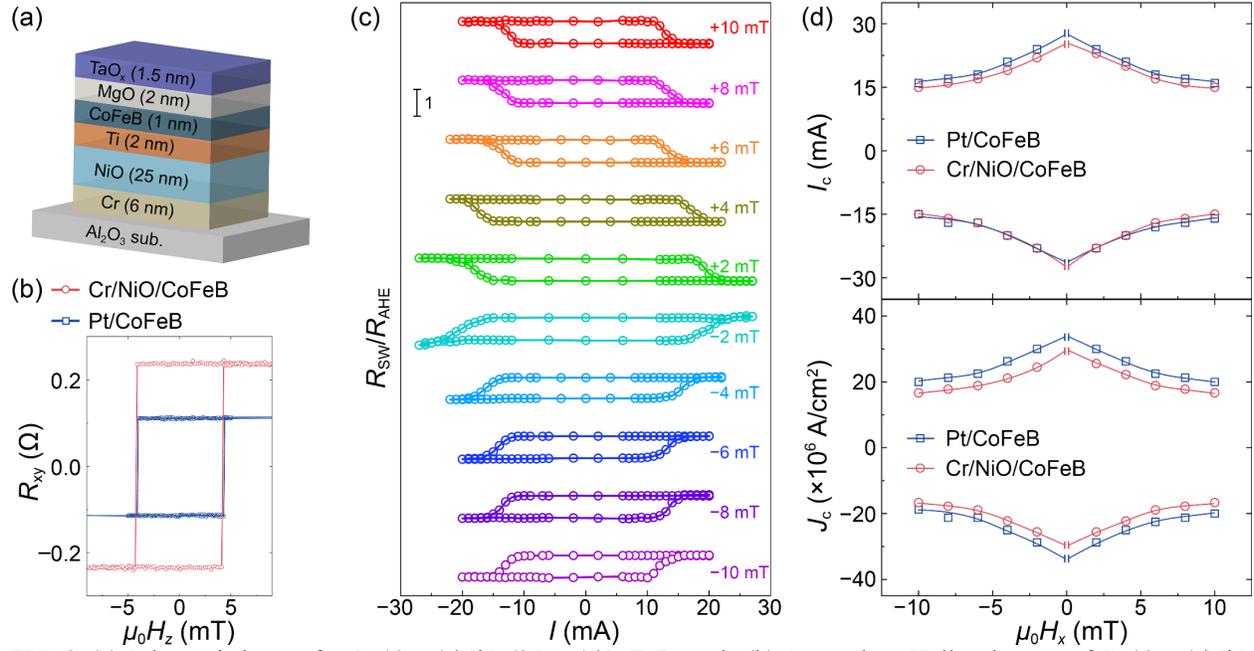

FIG. 3. (a) Schematic layout for Cr (6 nm)/NiO (25 nm)/CoFeB stack. (b) Anomalous Hall resistance of Cr (6 nm)/NiO (25 nm)/CoFeB and Pt (5 nm)/CoFeB with the out-of-plane magnetic field ($\mu_0H_z$). (c) The switching behaviors of Cr (6 nm)/NiO (25 nm)/CoFeB under various in-plane magnetic fields ($\mu_0H_x$). (d) The critical switching current ($I_c$) and critical current density ($J_c$) for Cr (6 nm)/NiO (25 nm)/CoFeB (the red line) and Pt (5 nm)/CoFeB (the blue line) under various $\mu_0H_x$.

To evaluate the contribution of the NiO/Py interface, we deposit NiO ($t$)/Py (6 nm) films on $Al_2O_3$ substrates and extract $\theta_T$ using in-plane SHH measurements. As shown in Fig. 4(a), $\theta_T$ increases with $t$ and saturates at −0.015 for $t \geq 5$ nm. Since NiO is an insulator and no Cr layer is present, the observed torque is attributed to spin torques generated at the NiO/Py interface [6] rather than long-range orbital torques [33,44]. When the NiO thickness reaches 5 nm, a well-defined NiO/Py interface is formed, leading to saturation of the torque efficiency for $t \geq 5$ nm. This weak interfacial spin torque indicates that the strong torque observed in the Cr (6 nm)/NiO (25 nm)/Py system is not attributed to the NiO/Py interface. Subsequently, we investigate the role of the Cr/NiO interface by inserting a 2.5 nm thick Cu layer between the Cr and NiO layers. Figure 4(b) presents the ST-FMR spectrum for the Cr (6 nm)/Cu (2.5 nm)/NiO (25 nm)/Py (6 nm) sample. The amplitude of $V_S$ is much smaller than $V_A$, with a ratio of $V_S/V_A = 0.11$, corresponding to $\theta_T = 0.026$. This value is significantly lower than that of the Cr/NiO/Py sample ($\theta_T = 0.097$). If the magnon torque were induced by the orbital Hall effect in Cr, a comparable magnon torque efficiency should be observed in this structure, since the orbital current generated in the Cr layer is expected to propagate through the Cu layer, leading to orbital accumulation at the Cu/NiO interface. This orbital accumulation is subsequently converted into spin accumulation through spin-orbit interaction in Ni, thereby exciting magnons in NiO and exerting torques on the Py layer. The observed significant reduction of $\theta_T$ in the Cr/Cu/NiO/Py system, therefore rules out the orbital Hall effect of Cr as the underlying mechanism.

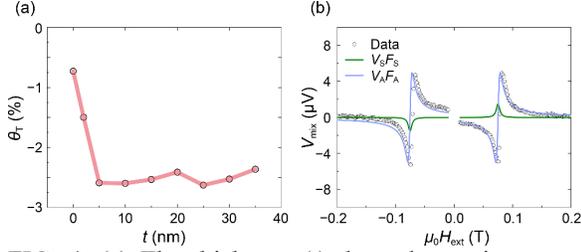

FIG. 4. (a) The thickness ($t$) dependent spin torque efficiency ($\theta_T$) of Al$_2$O$_3$/NiO ($t$)/Py (6 nm). The data are extracted from second harmonic Hall measurements. (b) ST-FMR spectra of Al$_2$O$_3$/Cr (6 nm)/Cu (2.5 nm)/NiO (25 nm)/Py.

## III. DISCUSSION

We discuss the role of the Cr/NiO interface in magnon torque generation. Previous studies have shown that orbital currents can be generated via the OREE at the metal/oxide interface, such as Cu/oxidized Cu [17,45] and Cu/Al$_2$O$_3$ [46]. However, this mechanism alone cannot explain the magnon torques observed in our Cr/NiO/FM system, since orbital angular momentum cannot couple directly to the magnetization through exchange interactions [27]. As a result, orbital currents cannot directly inject magnons into the NiO layer, and the observed magnon torques must be mediated by the spin accumulation. To further verify the origin of spin accumulation, an Al$_2$O$_3$/NiO (25 nm)/Cr (6 nm)/Py structure is fabricated and its $\theta_T$ is estimated to be −0.031 using SHH measurements (see Sec. VII of Supplemental Material [28]). This value is significantly lower than the $\theta_T$ = 0.101 observed in the Al$_2$O$_3$/Cr/NiO/Py sample, indicating the orbital Rashba-Edelstein effect dominates at the Cr/NiO interface and generates the orbital current, which is subsequently converted into spin accumulation. This spin accumulation can be understood in terms of inversion symmetry breaking at the Cr/NiO interface, which is induced by the orbital hybridization of Cr $t_{2g}$ $d$-orbitals through O $p$-orbitals [47] or Ni $d$-orbitals in the NiO layer. When the symmetry breaking energy is much stronger than the SOC, the Rashba spin splitting is maximized [48]. As a result, even though Cr possesses relatively weak SOC, a large Rashba spin splitting can be induced at the interface due to the strong inversion symmetry breaking. This Rashba spin splitting leads to spin accumulation at the Cr/NiO interface, which injects magnons into the NiO layer. Since the inversion symmetry breaking is along the $z$ direction, the orbital angular momentum is oriented in-plane. Consequently, the induced spin polarization is oriented along the $y$ direction, perpendicular to both the electric field ($x$-axis) and the symmetry-breaking direction ($z$-axis), in agreement with our experimental observations.

## IV. CONCLUSIONS

We experimentally demonstrate magnon torques in the Cr/NiO/FM heterostructure, providing evidence of orbital-to-magnon conversion. The magnon torque efficiency of the Cr/NiO/Py system is estimated to be 0.097 using ST-FMR techniques, which is significantly higher than that of the Cr/Py system (0.014). The corresponding effective spin Hall conductivity reaches 2.45 × 10$^5$ $\hbar/2e$ $\Omega^{-1}$ m$^{-1}$, surpassing that of the best conventional magnon torque system, PtTe$_2$/WTe$_2$/NiO/Py (1.08 × 10$^5$ $\hbar/2e$ $\Omega^{-1}$ m$^{-1}$), and comparable with electron-mediated spin source materials such as Pt (2.5 × 10$^5$ $\hbar/2e$ $\Omega^{-1}$ m$^{-1}$). We further realize magnon-torque-driven magnetization switching of a FM with PMA at room temperature. The corresponding switching power consumption density is as low as 0.136 mW μm$^{-2}$, comparable to that of Pt/CoFeB (0.113 mW μm$^{-2}$) (See Sec. V of Supplemental Material [28]). In addition, our experiments confirm that the magnon torques originate from the Cr/NiO interface, enabled by the inversion symmetry breaking due to the orbital hybridization between Cr and NiO. The magnon torques can be also observed in Cr/NiO/FM samples deposited on silicon wafers (see Sec. IX of Supplemental Material [28]), offering a promising route toward industrial applications.

## ACKNOWLEDGMENTS

This research is supported by National Research Foundation (NRF) Singapore Investigatorship (NRFI06-2020-0015) and the Ministry of Education, Singapore, under Tier 2 (MOE-T2EP50124-0006). [49-52]